# Publication Trend in an Indian Journal and a Pakistan Journal: A Comparative Analysis using Scientometric Approach

**M. Sadik Batcha**
Associate Professor
Department of Library and Information Science
Annamalai University, Tamil Nadu, India – 608002
msbau@rediffmail.com

**Muneer Ahmad**
Ph.D Research Scholar
Department of Library and Information Science
Annamalai University, Tamil Nadu, India – 608002
muneerbangroo@gmail.com

**Abstract**

*Scientometric analysis of 146 and 59 research articles published in Indian journal of Information Sources and Services (IJISS) and Pakistan Journal of Library and Information Science has been carried out. Seven Volumes of the IJISS containing 14 issues and Seven volumes of PJLIS containing 8 issues from 2011 – 2017 have been taken into consideration for the present study. The number of contributions, authorship pattern & author productivity, average citations, average length of articles, average keywords and collaborative papers has been analyzed. Out of 146 of IJISS contributions, only 39 are single authored and rest by multi authored with degree of collaboration 0.73 and week collaboration among the authors and from 59 contributions of PJLIS only 18 are single authored and rest by multi authored with degree of collaboration 0.69 and week collaboration among the authors. The study revealed that the author productivity is 0.53 (IJISS) and 0.50 (PJLIS) and dominated by the Indian and Pakistani authors.*

**Electronic access**
The journal is available at www.jalis.in

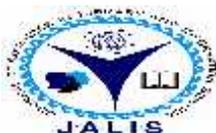

## 1.0 INTRODUCTION

Scientometrics is a discipline which analyses scientific publications to explore the structure and growth of science. The bibliometric or Scientometric or informetric techniques used to analyze various quantitative or qualitative aspects of publications. It is a scientific field that studies the evolution of science through some quantitative measures of scientific information, as the number of scientific articles published in a given period of time, their citation impact, etc. The history of science and technology, philosophy of science and sociology of scientific knowledge are the related fields of Scientometrics. The term scientometrics is often used with the meaning as the bibliometrics, originated in Russia. The application of quantitative methods to the history of science, scientometrics is the science of measuring the science, which involves counting artefacts to the production & use of information and arriving conclusions from the counts. Bibliometrics / Scientometrics research includes studies related to the scattering & growth of literature, author productivity, obsolescence of documents, distribution of scientific literature by country, by language, etc, which helps to monitor the growth & pattern of research.(Pritchard, 1969)[1] described the Bibliometrics as the application of mathematics and statistical methods to books and other media. Scientometric research is devoted to quantitative studies of science and technology (Van Raan, 1997)[2].Scientometrics applies the bibliometric techniques to science and examines the development of the sciences. (Virgil, 1994)[3]. Main areas of Scientometrics are individual scientific documents, authors, scientific institutions, academic journals and regional aspects of science (Stock & Sonja, 2006)[4]

In this paper, an attempt has been made to analyze the contributions to Indian Journal Information Sources and Services (IJISS) and Pakistan Journal of Library and Information Science (PJLIS) published during the year 2011 – 2017, in order to explore the author pattern, collaborative research, keywords and length of the papers among the contributions. This study covers the 146 and 59 articles of 14 and 8 issues published.

## 2.0 SOURCE OF STUDY

Indian Journal of Information Sources and Services (IJISS) and Pakistan Journal of Library and Information Science (PJLIS) were selected as the source journals for the present research study. Indian

Journal of Information Sources and Services (IJISS) is a popular journal of library and information science (LIS) in India. IJISS was started in 2010 as half yearly journal. With a gap of 8 years, fourteen issues were published up to 2017.Indian Journal of Information Sources and Services (IJISS) a Multidisciplinary Library and Information Science Journal is a refereed Journal and it publishes from India. This Journal encompasses all branches of Library and Information Sciences and its sub – disciplines like Library Management; Information Systems & Services, Information Processing & Retrieval; Information Sources & Services; Community Information System, Scientometrics & Informetrics, IR Theory, Knowledge Organization; Information processing & retrieval, Classification, Preservation & Conservation, Information Management, Library classification, Information sources, Systems and Services, Computer Application in Library; Digital Library; Information Systems; Bibliographic Control, etc. Pakistan Journal of Library and Information Science (PJLIS) is a popular journal of library and information science (LIS) in Pakistan. PJLIS, which was known as Pakistani Librarian till 1999, was started in 1995 as an annual journal by the Department of Library and Information Science, University of the Punjab, Lahore. With a gap of six years, eight issues were published up to 2016. Initially it was bilingual, but since volume 2006 it has been published in English language only. (Warraich & Ahmad, 2011)[5]

### 3.0 REVIEW OF RELATED LITERATURE

Scientometric Citation studies have done earlier by different authors on the different individual journal publications and literature on specific subject areas. The following studies related to the objectives of this study have been reviewed. (Srimurugan & Nattar, 2009)[6] analyzed the D-LIB magazine published during 2000 – 2007 which revealed that highest number of paper was published in 2005 and the lowest in 2007.(Vijay & Raghavan, 2007)[7] analyzed the Journal of Food Science & Technology published during 2000 – 2004 and found that above 93% of contributions were by multiple authors.A Scientimetric Analysis on Indian Journal of Physics was made by (Nattar,2009 )[8] during 2004 – 2008 which revealed that the year 2004 records the highest % of contributions regarding single, two and three authored. Kannappanavar B U, Swamy C & Vijay Kumar M (2004)[9] analyzed the publishing trends of Indian Chemical Scientists during 1996 – 2000, which revealed average number of authors per paper has increased from 7.52 to 8.39. An attempt was made by Shabana Tabusum (2003)[10] to analyze the Digital Literacy awareness of users in which focussed on the importance of Scientometrics for maximal use of sources. Guan & Ma (2007)[11]examined the China's Semiconductor Literature and found mega authored papers records the higher value for Co-Authorship Index. Tabusum S (2010)[12] analyzed the issues of Library Automation which have directly impacted on the scientometics studies which revealed that maximum number of papers were used according to category. A bibliometric study has been carried out by Kalyane V L and Sen B K (1995)[13] on the Journal of Oilseeds Research published during 1984 – 1992 which revealed that the keyword "Groundnut" tops the list with 53 records. Sanni S A and Zainab A N (2010)[14]examined the contributions published in Medical Journal of Malaysia during 2004 – 2008 and found 4.82% (28) of contributions were published by Malaysian authors with foreign collaboration.

### 4.0 OBJECTIVES OF THE STUDY

The following objectives of this study are formulated for the purpose of present study
- To map the year wise distribution of papers
- To examine the authorship pattern & author productivity
- To determine the degree of collaboration
- To assess the pattern of Co-Authorship
- To identify collaborative pattern
- To find the average length of papers
- To find the average keywords

### 5.0 SCOPE AND METHODOLOGY

The present study tries to find out the literature growth, authorship and collaboration pattern, average length of articles and average keywords included in the source journals. Seven Volumes of Indian Journal of Information Sources and Services (IJISS) and Pakistan Journal of Library and Information Science (PJLIS), published between 2011 and 2017 containing 14 and 8 issues have been taken into consideration to the present study. A datasheet was prepared in MS-Excel to record the data and then the data was entered manually into it from the journal itself. The details regarding number of papers, nature of author, keywords and length of papers are collected to fulfill the objectives of the present study. The collected data was analyzed with the following bibliometric indicators.
- Extent of Authorship Pattern (Single vs. Multiple)

- Degree of Collaboration
- Co-Authorship Index

## 6.0 LIMITATIONS

- The study includes open source journal in particular Indian Journal of Information Sources and Services (IJISS) and Pakistan Journal of Library and Information Science (PJLIS) among other journals.
- This study is limited to research papers published in Indian Journal of Information Sources and Services (IJISS) and Pakistan Journal of Library and Information Science (PJLIS) between the periods 2011 and 2017 only.

## 7.0 ANALYSIS AND DISCUSSION

### 7.1 Year wise Distribution of Papers

Table 1 shows the distribution of research articles published in IJISS and PJLIS during 2011 – 2017. The total of 146 and 59 research articles were published with an average of 20.86 and 9.83 articles per year. Out of 146 articles of IJISS, the highest number of research articles were published in the year 2012 with 29 research articles (14.5 articles per issue) followed by the 2011 ; 27 articles, 2014 ; 25 articles, 2017 ; 19 articles and the lowest number of articles were published in the year 2015 and 2016 with 14 articles each (7 articles per issue). Out of 59 articles of PJLIS, the highest number of research articles were published in the year 2016 with 30 research articles (10 articles per issue) followed by the 2012-2015 ; 6 articles each year, and the lowest number of articles were published in the year 2011 with 5 articles (8.47 articles per issue).

**Table 1:** Year Wise Distribution of Papers

| Year | Vol.No. | | No. of Issues | | Total Papers | | Percentage | | C.No. of Papers | | % of Cum. total | |
|---|---|---|---|---|---|---|---|---|---|---|---|---|
| | IJISS | PJLIS | IJISS | PJLIS | IJISS | PJLIS | IJISS | PJLIS | IJISS | PJLIS | IJISS | PJLIS |
| 2011 | 1 | 12 | 2 | 1 | 27 | 5 | 18.49 | 8.47 | 27 | 5 | 18.49 | 8.47 |
| 2012 | 2 | 13 | 2 | 1 | 29 | 6 | 19.86 | 10.17 | 56 | 11 | 38.36 | 18.64 |
| 2013 | 3 | 14 | 2 | 1 | 18 | 6 | 12.33 | 10.07 | 74 | 17 | 50.68 | 28.81 |
| 2014 | 4 | 15 | 2 | 1 | 25 | 6 | 17.12 | 10.17 | 99 | 23 | 67.81 | 38.98 |
| 2015 | 5 | 16 | 2 | 1 | 14 | 6 | 9.59 | 10.17 | 113 | 29 | 77.40 | 49.15 |
| 2016 | 6 | 17 | 2 | 1 | 14 | 20 | 9.59 | 33.90 | 127 | 49 | 86.99 | 83.05 |
| 2016,17 | 7 | 18 | 2 | 2 | 19 | 10 | 13.02 | 16.95 | 146 | 59 | 100 | 100 |
| Total | | | 14 | 8 | 146 | 59 | 100 | 100 | | | | |

### 7.2 Authorship Pattern

The table 2 presents the data according to rank. As for as IJISS is concerned it is observed from the Table 2,about 73% of papers were contributed by multi authors. Out of 146 papers, the highest numbers of papers were published by double authors and it accounts for 84 with 57.53% followed by single authored articles account for 39 with 26.71%. 14.38% of articles were published by three authors. 1.37 % of articles were published by four & above authors. But the trend of the author pattern in the journal shows that the team size was two to three. For PJLIS, about 69% of papers were contributed by multi authors. Out of 59 papers, the highest number of papers was published by double authors and it accounts for 26 with 44.07% followed by single authored articles account for 18 with 30.51%. 20.34% of articles were published by three authors. 5.08 % of articles were published by four authors. But the trend of the author pattern in the journal shows that the team size was two to three.

**Table 2 :** Authorship Pattern

| Rank | Authors | No. of Papers | | % | | Cum no. of Papers | | Cum % | |
|---|---|---|---|---|---|---|---|---|---|
| | | IJISS | PJLIS | IJISS | PJLIS | IJISS | PJLIS | IJISS | PJLIS |
| 1 | Two | 84 | 26 | 57.53 | 44.07 | 84 | 26 | 57.53 | 44.07 |
| 2 | Single | 39 | 18 | 26.71 | 30.51 | 123 | 44 | 84.25 | 74.58 |
| 3 | Three | 21 | 12 | 14.38 | 20.34 | 144 | 56 | 98.63 | 94.92 |
| 4 | Four & Above | 2 | 3 | 1.37 | 5.08 | 146 | 59 | 100 | 100 |
| | **Total** | **146** | **59** | **100%** | **100%** | | | | |

**7.3 Authorship Pattern in accordance with year wise**

The data pertaining to authorship pattern year wise have been given in the Table No.3. Regarding single authored contributions, the years 2012 & 2011 have the highest contributions with 12 and 8 each and 6 in year 2015 and 4 in the year 2014 & 2017 and lowest 2 in the year 2016. Regarding double authored contributions, the year 2014 has the highest contributions with 17. The year 2011 has the highest contributions regarding three authored contributions with 6 and 2012. The year 2011 & 2014 has the highest contributions of four authored (more than three authors) with 1each. PJLIS data regarding single authored contributions, the years 2016 & 2012 have the highest contributions with 10 and 4 each and the zero in year 2015. Regarding double authored contributions, the year 2016 has the highest contributions with 14. The year 2016 has the highest contributions regarding three authored contributions with 5 and 2015 with 2. The year 2015 has the highest contributions of four authored (more than three authors)

**Table 3: Authorship** Pattern - year wise

| Year | Authors | | | | | | | | | |
|---|---|---|---|---|---|---|---|---|---|---|
| | 1 | | 2 | | 3 | | 4 | | Total | |
| | IJISS | PJLIS | IJISS | PJLIS | IJISS | PJLIS | IJISS | PJLIS | IJISS | PJLIS |
| 2011 | 8 | 2 | 12 | 3 | 6 | 0 | 1 | 0 | 27 | 5 |
| 2012 | 12 | 4 | 14 | 0 | 3 | 2 | 0 | 0 | 29 | 6 |
| 2013 | 3 | 1 | 13 | 4 | 2 | 1 | 0 | 0 | 18 | 6 |
| 2014 | 4 | 1 | 17 | 3 | 3 | 2 | 1 | 0 | 25 | 6 |
| 2015 | 6 | 0 | 6 | 2 | 2 | 2 | 0 | 2 | 14 | 6 |
| 2016 | 2 | 10 | 10 | 14 | 2 | 5 | 0 | 1 | 14 | 30 |
| 2017 | 4 | - | 12 | - | 3 | - | 0 | - | 19 | - |
| Total | 39 | 18 | 84 | 26 | 21 | 12 | 2 | 3 | 146 | 59 |

**7.4 Author Productivity**

The data pertaining to author productivity has presented in the Table 4. The table for IJISS shows that the total average number of authors per paper is 1.89 for the 146 articles. The years 2013 & 2017 has the relatively equal average number of authors per article when compared the total average number of authors per article. The average productivity per author is 0.53 during the year 2011 - 2017. The years 2013 & 2017 has the relatively equal productivity per author when compared to the average productivity.

The table for PJLIS shows that the total average number of authors per paper is 1.98 for the 59 articles. The years 2011 & 2012 has the relatively equal average number of authors per article when compared the total average number of authors per article. The average productivity per author is 0.50 during the year 2011 - 2016. The years 2016 has the relatively equal productivity per author when compared to the average productivity. Productivity has been calculated with the following formula (Fuyuki, 2009)[15] Average Authors per Paper = No. of Authors / No. of Papers .Productivity per Author = No. of Papers / No. of Authors

**Table 4:** Author Productivity – year wise

| S.No. | Year | Total No. of Papers | | Total No. of Authors | | AAPP | | Productivityper Author | |
|---|---|---|---|---|---|---|---|---|---|
| | | IJISS | PJLIS | IJISS | PJLIS | IJISS | PJLIS | IJISS | PJLIS |
| 1 | 2011 | 27 | 5 | 54 | 8 | 2 | 1.6 | 0.5 | 0.63 |
| 2 | 2012 | 29 | 6 | 47 | 9 | 1.62 | 1.5 | 0.62 | 0.67 |
| 3 | 2013 | 18 | 6 | 35 | 12 | 1.94 | 2 | 0.51 | 0.5 |
| 4 | 2014 | 25 | 6 | 51 | 13 | 2.04 | 2.17 | 0.49 | 0.46 |
| 5 | 2015 | 14 | 6 | 24 | 18 | 1.71 | 3 | 0.58 | 0.33 |
| 6 | 2016 | 14 | 30 | 28 | 57 | 2 | 1.9 | 0.5 | 0.53 |
| 7 | 2017 | 19 | - | 37 | - | 1.95 | - | 0.51 | - |
| | Total | 146 | 59 | 276 | 117 | 1.89 | 1.98 | 0.53 | 0.50 |

**7.5 Degree of Collaboration**
In order to determine the strength of Collaboration (DC), the following formula suggested by (Subramanyam, 1993)[16] has been employed.
DC = $N_m/N_m+N_s$, Where, DC = Degree of Collaboration, Nm = Number of Multiple Authored Papers, Ns = Number of Single Authored Papers

The Degree of Collaboration of authors by year wise is presented in the Table 5. The degree of collaboration for IJISS ranges from 0.74 to 0.93. The average degree of collaboration is 0.73 during the period 2011 – 2017 and it brings out clearly that there exists a high level of collaboration in the journal. The degree of collaboration for PJLIS ranges from 0.33 to 1. The average degree of collaboration is 0.69 during the period 2011 – 2016 and it brings out clearly that there exists a low level of collaboration in the journal.

**Table 5**: Degree of Collaboration – year wise

| S.No. | Year | $N_s$ | | $N_m$ | | $N_s + N_m$ | | DC | |
|---|---|---|---|---|---|---|---|---|---|
| | | IJISS | PJLIS | IJISS | PJLIS | IJISS | PJLIS | IJISS | PJLIS |
| 1 | 2011 | 8 | 2 | 46 | 3 | 54 | 5 | 0.85 | 0.6 |
| 2 | 2012 | 12 | 4 | 35 | 2 | 47 | 6 | 0.74 | 0.33 |
| 3 | 2013 | 3 | 1 | 32 | 5 | 35 | 6 | 0.91 | 0.83 |
| 4 | 2014 | 4 | 1 | 47 | 5 | 51 | 6 | 0.92 | 0.83 |
| 5 | 2015 | 6 | 0 | 18 | 6 | 24 | 6 | 0.75 | 1 |
| 6 | 2016 | 2 | 10 | 26 | 20 | 28 | 30 | 0.93 | 0.67 |
| 7 | 2017 | 4 | - | 33 | - | 37 | - | 0.89 | - |
| | Total | 39 | 18 | 107 | 41 | 146 | 59 | 0.73 | 0.69 |

**7.6 Pattern of Co-Authorship**
In order to assess the Pattern of Co-Authorship (CAI), the following formula suggested by (Garg and Padhi, 1999)[17] has been employed.
Where,
Nij = Number of papers having authors in block i
Nio = Total output of block i
Noj = Number of papers having j authors for all blocks
Noo = Total number of papers for all authors and all blocks

CAI = 100 implies that a country's co-authorship effort for a particular type of authorship corresponds to the world average, CAI > 100 reflects higher than average co-authorship effort, and CAI < 100 lower than average co-authorship effort by that country for a given type of authorship pattern.

For calculating the co-authorship index for authors, countries have been replaced by block. For this study, the authors have been classified into four blocks, viz Single, Two, Three and more than three authors and the results of Co-authorship index as per the formula have been presented in the Table No.6(a) and 6(b).

**Table 6 (a)**: Pattern of Co-Authorship (IJISS) - year wise

| S.No. | Year | Single Author | | Two Authors | | Three Authors | | >Three Authors | | Total |
|---|---|---|---|---|---|---|---|---|---|---|
| | | No | CAI | No | CAI | No | CAI | No | CAI | |
| 1 | 2011 | 8 | 55 | 12 | 38 | 6 | 77 | 1 | 135 | 27 |
| 2 | 2012 | 12 | 95 | 14 | 51 | 3 | 44 | 0 | 0 | 29 |
| 3 | 2013 | 3 | 33 | 13 | 66 | 2 | 40 | 0 | 0 | 18 |
| 4 | 2014 | 4 | 29 | 17 | 57 | 3 | 41 | 1 | 143 | 26 |
| 5 | 2015 | 6 | 93 | 6 | 43 | 2 | 57 | 0 | 0 | 14 |
| 6 | 2016 | 2 | 26 | 2 | 12 | 2 | 49 | 0 | 0 | 6 |
| 7 | 2017 | 4 | 40 | 12 | 56 | 3 | 56 | 0 | 0 | 19 |
| | Total | 39 | | 84 | | 21 | | 2 | | 146 |

**Table 6 (b):** Pattern of Co-Authorship (PJLIS) - year wise

| S.No. | Year | Single Author | | Two Authors | | Three Authors | | >Three Authors | | Total |
|---|---|---|---|---|---|---|---|---|---|---|
| | | No | CAI | No | CAI | No | CAI | No | CAI | |
| 1 | 2011 | 2 | 131 | 3 | 136 | 0 | 0 | 0 | 0 | 5 |
| 2 | 2012 | 4 | 218 | 0 | 0 | 2 | 164 | 0 | 0 | 6 |
| 3 | 2013 | 1 | 55 | 4 | 151 | 1 | 82 | 0 | 0 | 6 |
| 4 | 2014 | 1 | 55 | 3 | 113 | 2 | 164 | 0 | 0 | 6 |
| 5 | 2015 | 0 | 0 | 2 | 76 | 2 | 164 | 2 | 656 | 6 |
| 6 | 2016 | 10 | 109 | 14 | 106 | 5 | 82 | 1 | 66 | 30 |
| Total | | 18 | | 26 | | 12 | | 3 | | 59 |

As for as IJISS is concerned it is observed from the Table 6(a), the CAI for single authors is declined from 55 in the year 2011 to 40 in the year 2017. On the other hand, the CAI for double authors is enhanced from 38 in the year 2011 to 56 in the year 2017, which indicates the pattern of co authorship is increasing among the contributions of the journal. On the other hand, there is a fluctuation trend of CAI for multi authored contributions. The similar type of result has been drawn by Jeyshankar R, et al in the Current Science. (Jeyshakar, 2009)[18.] It is observed from the Table 6(b), the CAI for single authors is declined from 131 in the year 2011 to 109 in the year 2016. In the same way, the CAI for double authors is also declined from 136 in the year 2011 to 106 in the year 2016, which indicates the pattern of co-authorship is decreasing among the contributions of the journal. On the other hand, there is a fluctuation trend of CAI for multi authored contributions.

### 7.7 Distribution of Pages

Table 7 shows that 146 papers published with a total page of 817 (average 5.60 pages per article) during the year 2011 – 2017. It is observed that the average length of the articles varied from a minimum of 4.41 pages to a maximum of 11.67 pages. The year 2016 has highest average page per paper with 7.07 pages while the year 2012 has the lowest average page per paper with 4.41.And for PJLIS it shows that 59 papers published with a total page of 593 (average 10.05 pages per article) during the year 2011 – 2016. It is observed that the average length of the articles varied from a minimum of 6.83 pages to a maximum of 11.67 pages. The year 2016 has highest average page per paper with 11.67 pages while the year 2013 has the lowest average page per paper with 6.83.

**Table 7 – Distribution of Pages - year wise**

| S.No. | Year | No. Of Articles | | Total Pages | | Average Pages Per Article | | Percentage | | Cum No. of Pages | | Cum % | |
|---|---|---|---|---|---|---|---|---|---|---|---|---|---|
| | | IJISS | PJLIS | IJISS | PJLIS | IJISS | PJLIS | IJISS | PJLIS | IJISS | PJLIS | IJISS | PJLIS |
| 1 | 2011 | 27 | 5 | 154 | 47 | 5.7 | 9.4 | 18.85 | 8.47 | 154 | 47 | 18.85 | 7.93 |
| 2 | 2012 | 29 | 6 | 128 | 48 | 4.41 | 8.00 | 15.67 | 10.17 | 282 | 95 | 34.52 | 16.02 |
| 3 | 2013 | 18 | 6 | 97 | 41 | 5.39 | 6.83 | 11.87 | 10.17 | 379 | 136 | 46.39 | 22.93 |
| 4 | 2014 | 25 | 6 | 145 | 53 | 5.8 | 8.83 | 17.75 | 10.17 | 524 | 189 | 64.14 | 31.87 |
| 5 | 2015 | 14 | 6 | 99 | 54 | 7.07 | 9.00 | 12.12 | 10.17 | 623 | 243 | 76.25 | 40.98 |
| 6 | 2016 | 14 | 30 | 82 | 350 | 5.86 | 11.67 | 10.04 | 50.85 | 705 | 593 | 86.29 | 100 |
| 7 | 2017 | 19 | - | 112 | - | 5.89 | - | 13.71 | - | 817 | - | 100.00 | - |
| Total | | 146 | 59 | 817 | 593 | 5.60 | 10.05 | 100 | 100 | | | | |

### 7.8 Average Keywords per Article

Table 8 reveals that for IJISS 581 keywords have been appended to 146 papers. It is observed that the average keyword of the paper varied from a minimum of 3.14 to a maximum of 4.53 during the year 2011 – 2017. The year 2017 has the highest average keyword per paper with 4.53 keywords per paper while the year 2012 has the lowest average

keywords per paper with 3.14. The overall average keywords per article are 3.98. Table 8 reveals that 268 keywords have been appended to 59 papers. And it is observed that in PJLIS the average keyword of the paper varied from a minimum of 3 to a maximum of 5.33 during the year 2011 – 2016. The year 2014 has the highest average keyword per paper with 5.33 keywords per paper while the year 2012 has the lowest average keywords per paper with 3. The overall average keywords per article are 4.54.

**Table 8:** Average Keywords per Article – Year wise

| S. No. | Year | No. of Articles | | Total Keywords | | Average Keywords per Paper | |
|---|---|---|---|---|---|---|---|
| | | IJISS | PJLIS | IJISS | PJLIS | IJISS | PJLIS |
| 1 | 2011 | 27 | 5 | 97 | 19 | 3.59 | 3.8 |
| 2 | 2012 | 29 | 6 | 91 | 18 | 3.14 | 3 |
| 3 | 2013 | 18 | 6 | 80 | 25 | 4.44 | 4.17 |
| 4 | 2014 | 25 | 6 | 110 | 32 | 4.4 | 5.33 |
| 5 | 2015 | 14 | 6 | 59 | 24 | 4.21 | 4 |
| 6 | 2016 | 14 | 30 | 58 | 150 | 4.14 | 5 |
| 7 | 2017 | 19 | - | 86 | - | 4.53 | - |
| | Total | 146 | 59 | 581 | 268 | 3.98 | 4.54 |

**7.9 Distribution of Indian and Foreign Contributions**

Table 9(a) and 9(b) shows that out of 146 articles, 138 (94.52%) articles published by Indian authors followed by International authors with 6 articles (4.11%). Only 2 (1.37%) articles published by Indian Authors collaborated with international authors and for PJLIS out of 59 articles, 94 (77.05%) articles published by Pakistani authors followed by International authors with 23 articles (18.85%). Only 5 (4.10%) articles published by Pakistani Authors collaborated with international authors and similar type of study has been conducted by (Zainab A N, et al, 2009)[19]. It seems that there was poor collaboration of Indian and Pakistani authors with Foreign Authors. It is observed from the data that out of 14 issues of IJISS, 8 issues having the contributions only by Foreign Authors.
And from the data of PJLIS it is observed that out of 8 issues, 5 issues having the contributions only by Foreign Authors.

**Table 9(a):** Distribution of Indian and Foreign Contributions

| Form | Contributions | % |
|---|---|---|
| Indian Authors | 138 | 94.52 |
| Indian Authors with ForeignCollaboration | 2 | 1.37 |
| Foreign Authors | 6 | 4.11 |
| Total | 146 | |

**Table 9(b):** Distribution of Pakistani and Foreign Contributions

| Form | Contributions | % |
|---|---|---|
| Pakistani Authors | 94 | 77.05 |
| Pakistani Authors with Foreign Collaboration | 5 | 4.10 |
| Foreign Authors | 23 | 18.85 |
| Total | 122 | |

**7.0 Findings and Conclusion**

The study has revealed the findings which will be useful to information managers and persons associated with Indian Journal of Information Sources and Services (IJISS) and Pakistan Journal of Library and Information Science.In IJISS the maximum number of papers published in the year 2012 and minimum in the years 2015 & 2016.The highest number of research papers found in the study by multiple authors during the study period. The degree of collaboration found was 0.73.It is found that the average value for CAI was around 40 during the study period and it reflects below than the world average. It is an alarming state that effort should be made on par with world average. The author productivity is 0.53 and the average number of authors per paper found in the present study is 1.89.The average pages per paper observed in the study is 5.60 as the total numbers of pages are 817 with 146 papers. The average keywords per paper observed in the study are 3.98 as the total number of keywords are 581.The majority of the contributions are by Indian Authors (94.52%).Papers by Indian Authors with Foreign Collaboration are minimal (1.37% of articles).

The findings have focused that the majority of papers by multi authors and Indian authors. There was poor international collaboration by Indian authors. The average page is 5.60 and it is the ideal for research papers. The Degree of collaboration (using Subramanyam's formula) indicates that there exists a high degree of collaboration. The average Co-Authorship Index for all the authors reflects the world average in the journal and improving trend of co-authored papers. The study revealed that the journal seems to be not popular among the international research community with around 4.11% of papers.

As for as Pakistan Journal of Library and Information Science is conserned the maximum number of papers published in 2016 and minimum in 2011.The highest number of research papers contributed by multiple authors during the study period. The degree of collaboration was 0.69. It is found that the average value for CAI was around 100 during the study period and it reflects the world average. The author productivity is 0.50 and the average number of authors per paper is 1.98.The average pages per paper are 10.05. The average keywords per paper are 4.54. The majority of the contributions are by Pakistani Authors (77.05%). Papers by Pakistani Authors with Foreign Collaboration are minimal (4.10% of articles).The analysis explores that the majority of papers by multi authors and Pakistani authors. There was poor international collaboration by Pakistani authors. The average page is 10.05 and it is the ideal for research papers. The Degree of collaboration (using Subramanyam's formula) indicates that there exists a high degree of collaboration. The average Co-Authorship Index for all the authors reflects the world average in the journal and improving trend of co-authored papers. The study revealed that the journal seems to be popular among the international research community with around 19% of papers.